\documentclass[a4paper]{jpconf}

\usepackage{graphicx}

\begin{document}

\title{Baryonium Dark Matter: Events with Circle Configurations in the Atmosphere and on the Surface of Earth}

\author{Olga I. Piskounova}
\address{Senior Scientist, HE Particle Physics, P.N. Lebedev Physical Institute of RAS,119991  Moscow, Russia}

\ead{piskounovaoi@lebedev.ru}

\begin{abstract}

This research supposes the parallel study of the manifestations of Baryonium Dark Matter (BDM) in astroparticle collisions in the atmosphere and among the footprints of the massive extraterrestrial objects that fall on the surface of Earth. These objects differ from stone asteroids and are seen as lightning bolides that are burning with a blue flame. Since BDM is a QCD state, the characteristics of BDM disintegration events can be analyzed with the same Quark-Gluon Strings Model (QGSM) as hadroproproduction data on colliders. It was concluded that the HE event, which was detected in the stratosphere in 1975, differs from the collision of the nucleus. The signatures of such events are 1) the circle distribution of heavy secondary hadrons with almost nothing inside the ring and 2) the presence of heavy secondary particles with a mass out of range of known hadrons. The circle will also be the signature of heavy BDM conglomerate that disintegrate in the atmosphere (or in space) for a few similar lower-mass BDMs. These signatures should be observed in super-heavy BDM collapses on the Earth. The traces of such events are seen as circles of pits or shafts in the ground rocks. There are at least two pieces of evidence of such holes: cenotes on the Yucatan peninsula (Mexico) and pits in Durrington Walls (England). The vertical pits, or wells (because sometimes they are filled with water) have a depth of 5 - 50 meters. These super-heavy BDM collapses differ from the meteorite crash, because they do not form the craters. The existence of these places around the globe indicates the rare likelihood of meeting the super-massive BDM in proximity to Earth. The Tunguska meteorite could also be a BDM collapse. Finally, some suggestions are made about the emission of Baryonium Dark Matter toroidal conglomerates with the jets from giant active Supermassive Black Holes (SMBH).

\end{abstract}

\section{Introduction into Baryonium Dark Matter}

\begin{figure}[h]
\begin{center}
\includegraphics[scale=0.37]{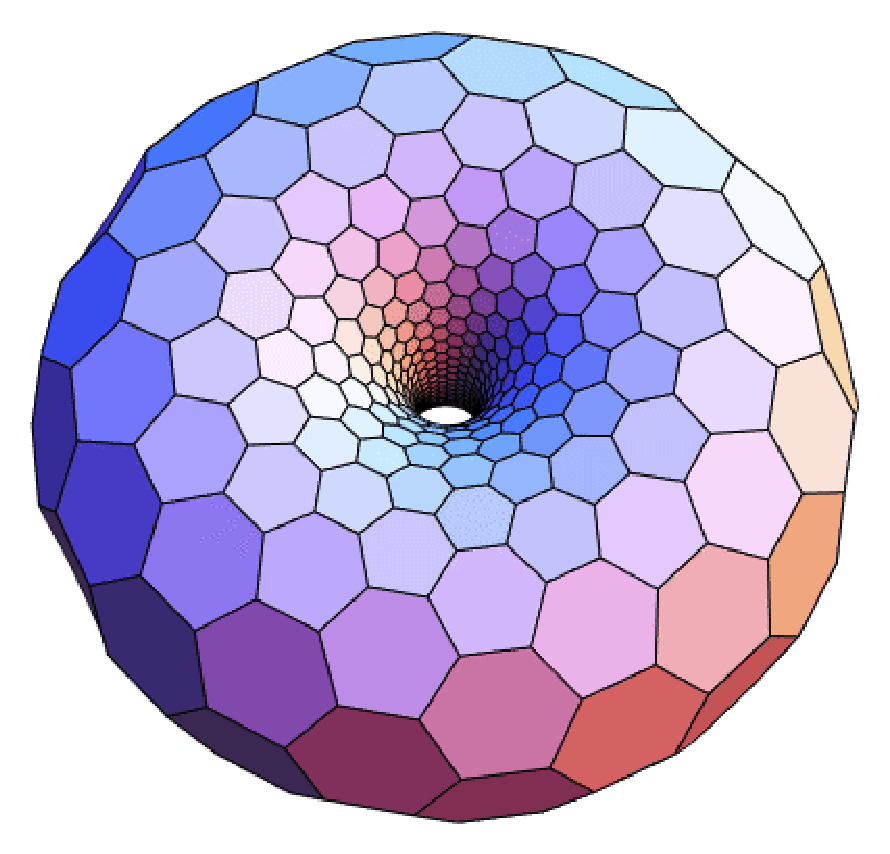}
\caption{ The torus that is covered with baryonium hexagons ( from collection of Wolfram community).} 
\label{torusWf}
\end{center}
\end{figure}

The asymmetry in baryon production at HE proton colliders tells us that proton string junction (SJ) brings positive baryon charge to the central rapidity region even at baryon collisions of high energy \cite{asymmetry}. It means that SJs play an important role in baryon interactions. Baryon and antibaryon string junctions can be connected into multiquark states: tetraquarks, pentaquarks, hexaquarks etc. The figure of quark-less bayonium states has been constructed in 3D form with completely self-connected SJs and anti-SJs. The complete connection of QCD diagrams is possible only on surface of torus\cite{torusWf}, see figure~\ref{torusWf}. 
 
This state of baryon-antibaryon matter has been named Baryonium Dark Matter (BDM) \cite{BDM} and should have discrete levels of masses \cite{masslevels}, because not every arbitrary number of baryon-antibaryon hexagons can form the torus. The tetraquarks pentaquarks,and hexaquarks includes the antiproton SJ, so can be considered as results of deconstruction of neutral BDM torus. 

In such a way, Dark Matter is an object of QCD string physics, which is responsible for baryon and antibaryon production on the wide logariphmic range of energies and masses begining from HE hadroproduction on LHC upto Ultra High Energy (UHE) matter
injected to space with the relativistic jets from SMBH.

\section{Stratosphere event of 1975}

Baryonium Dark Matter of medium masses can be seen in the cosmic ray physics. The high energy collision event was detected in stratosphere at 30 km hight in the calorimeter with lead layers and x-ray films \cite{stratosphere}. The electron-photon cascades have been developed in calorimeter and detected on x-ray films as dark spots. The spot on x film with the certain darkness gives us the energy of secondary particle. After the measurements we obtained the coordinates (x,y) and $E_{spot}$ of 106 particles for futher analysis. The rapidity of each particle was defined as:
$Y = - 0.5*\ln{((E-Pl)/(E+Pl))} = \ln{2E/(M_t)}$, where $M_t= \sqrt{Pt^2+M_0^2}$, E, Pl, Pt are energy and momenta of particle and $M_0$ - its mass. 

\begin{figure}[h]
\begin{center}
\includegraphics[scale=0.37]{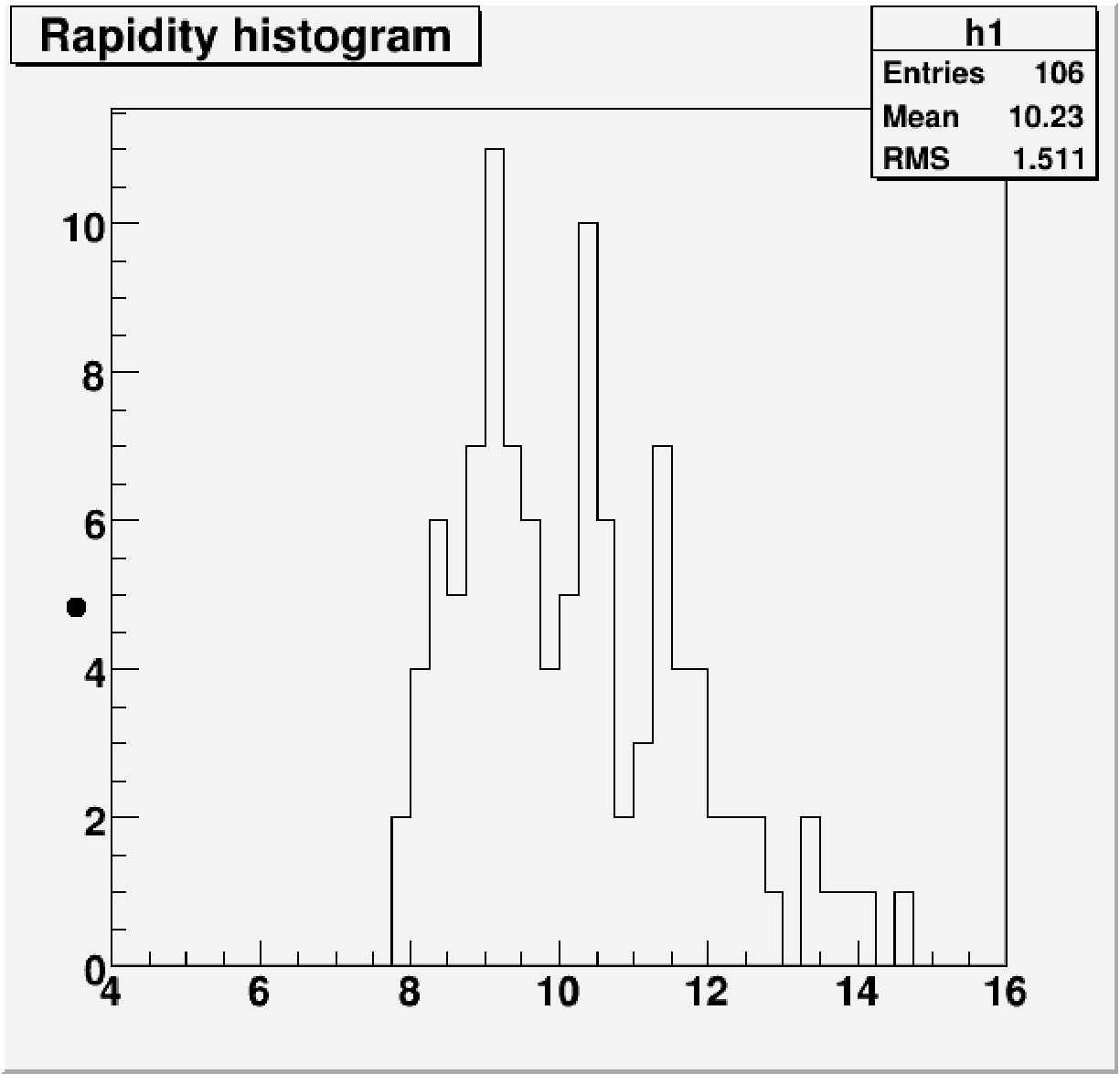}
\caption{ Rapidity histogram with peaks, which mean many secondary particles of the same mass are produced and detected withing the circle.}   
\label{Yhysto025}
\end{center}
\end{figure}

\begin{figure}[h]
\begin{center}
\includegraphics[scale=0.37]{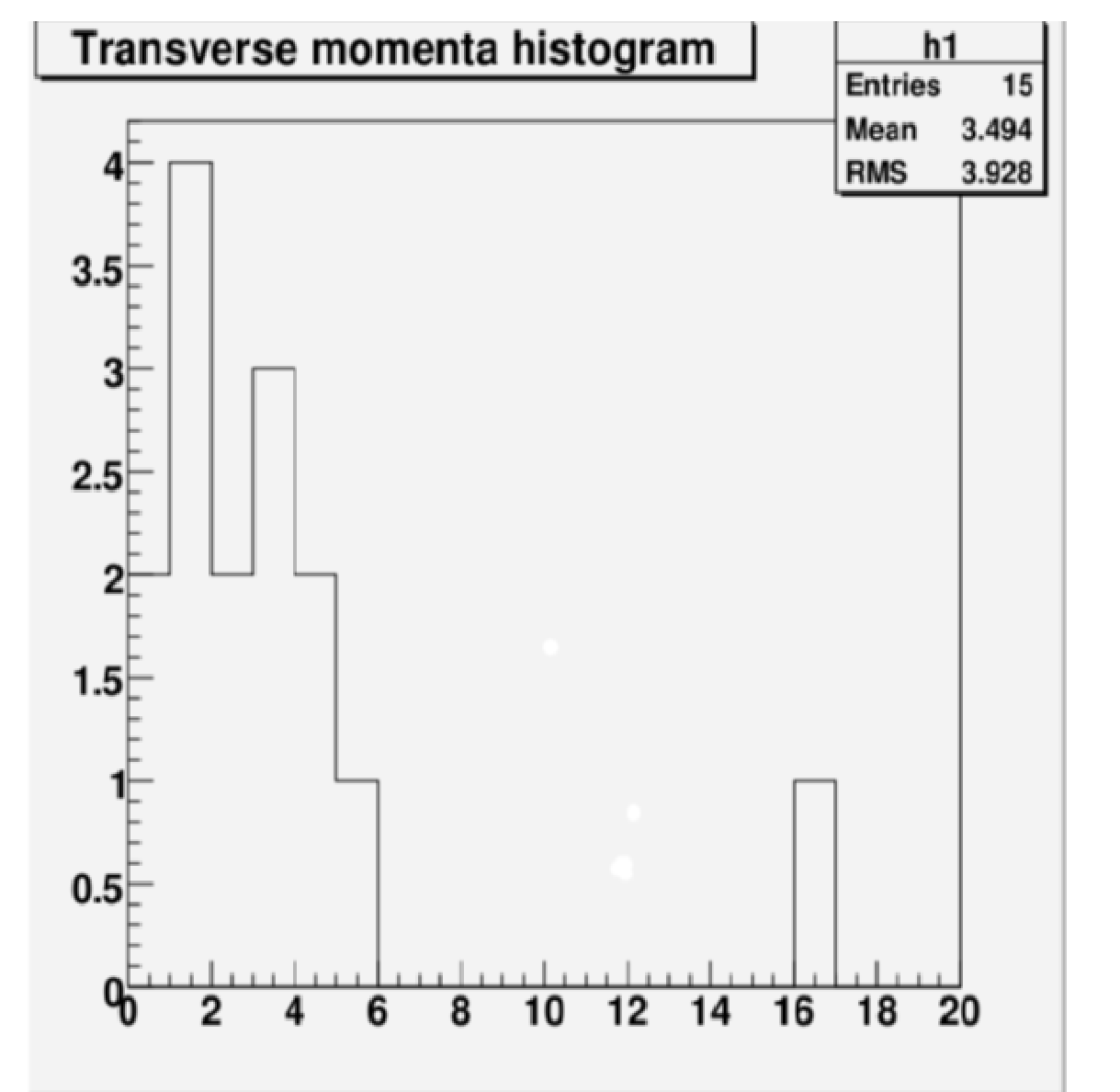}
\caption{ Transverse mass distribution for the central rapidity area, Y$\le 2$.}
\label{Mthystoa}
\end{center}
\end{figure}

As the result, the event data were histogramed  in rapidities and transverse masses, see figure~\ref{Yhysto025}, figure~\ref{Mthystoa}.

The rapidity histogram shows that maxmal rapidity for this collision is approximately $Y_{max}$ = 7.5 in the center-of-mass system. We could extract the energy of collision per one proton due to comparison with the results of ALICE nucleus-nucleus experiment at LHC \cite{alice}. $E_{c.m.s.}$ is approximatelly equal to 600 GeV. Average multiplicity per proton for this energy is 5 hadrons in central region of rapidity. If the histogram shows at least 28 hadrons, the projectile nucleus was carbon or a more heavy nucleus. But the transverse mass distributions that have been obtained for two part of spectra (for the central area and for fragmentation region) tells an opposite. 

The histogram in fragmentation region reports that the projectile was lightest nucleus with 2 protons \cite{kotelnikov}. In addition, the transverse mass histogram for central rapidity area, figure~\ref{Mthystoa}, demonstrates an unusualy heavy hadron with transverse mass more that 16 GeV. In such a way we have the procedure of analysis of the traces after BDM recoils wherever they happen.

These two peculiarities of event, together with the rapidity distribution in figure~\ref{Yhysto025}, which is showing the peaks at high rapidities, help us conclude that it was not a colliding nucleus. So, we have specific signatures for the collision of astroparticle, which could be supposed the Baryonium Dark Matter (BDM). The interaction of such particle gives multiple secondaries in central region and, what is most important, the circles at imtermediate rapidity, which may correspond to disintegration of BDM particle into identical constituents of lower masses. The area of maximal rapidity is almost empty of produced protons, so the projectile is not a nucleus. The preliminary analysis of the event has already discovered such circles \cite{dremin}, but author did not relate them with heavy BDM disintegration.

\section{Footprints of BDM collapse events on the surface of the Earth}

\begin{figure}[h]
\begin{center}
\includegraphics[scale=0.37]{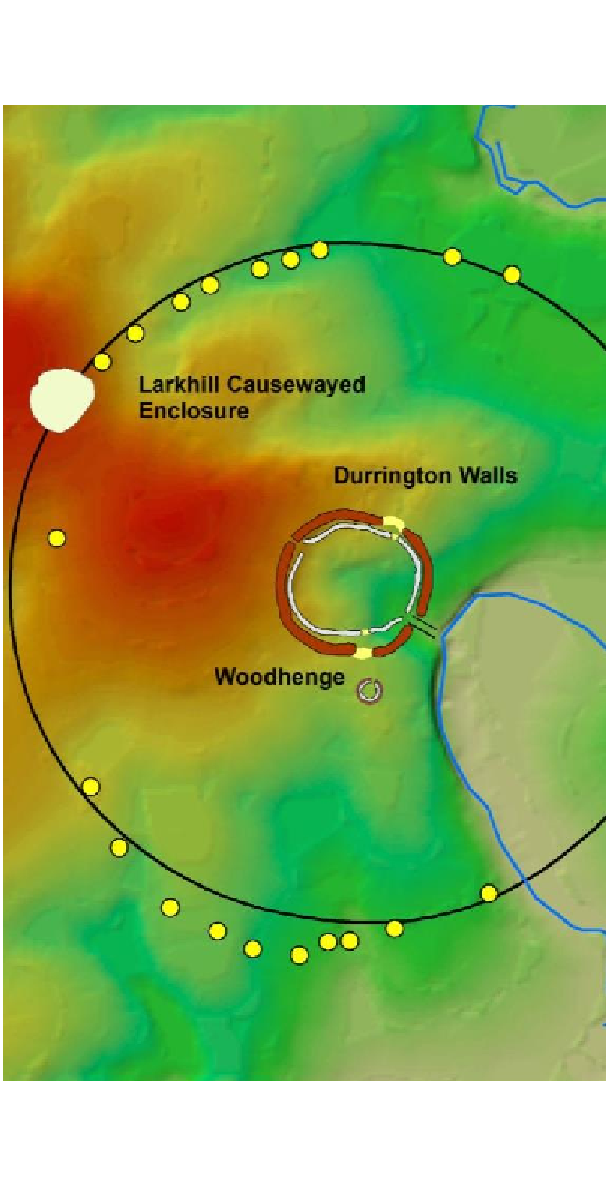}
\caption{The circle of pits on the ground at Durrington Walls. Originaly, the pits (or wells) may be filled with water and have been revered as sacral place.}
\label{DW}
\end{center}
\end{figure}

\begin{figure}[h]
\begin{center}
\includegraphics[scale=0.37]{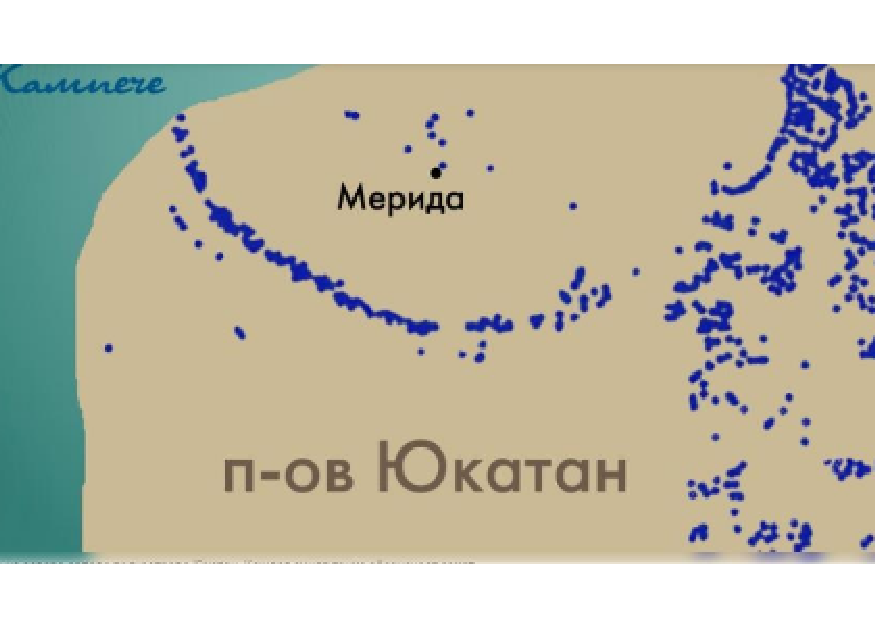}
\caption{The circle of cenots on the Yucatan peninsula.} 
\label{yucatan}
\end{center}
\end{figure}

The supermassive Baryonium Dark Matter conglomerates could certainly reach the surface of Earth. What are we expecting to see?
As it was found from the views of two locations with misterious circles, see figure~\ref{DW} and figure~\ref{yucatan}: pits in Durrington Walls \cite{DW} and cenotes on Yucatan \cite{Yucatan} peninsula have been organized in circles. The central areas of circles are almost empty, because, as it was seen from histograms of stratosphere event, there should be UHE light components: electromagnetic radiation, hadron showers and light components of BDM, which can not destroy the underground rock. The circle of pits is the signature of the first stage disintegration  of supermassive Baryonium Dark Matter conglomerate for the number of BDMs of equal lower masses. What is important, there are no craters, as for the falls of meteorites. As it was seen sometimes in our time, the fall of Baryonium Dark Matter is looking as a flight of  lightening bolide in the sky.

There is a courious question arising, why the ring in Durrigton Walls could not be built by humans? It is clearly seen that the circle of pits is not ideal, there are some fluctuations in the positions of pits. Humans always have ideal plans in minds, before to realize them, like pyramids etc. From this point of view, the pits in Durrington Walls are not forming ideal circle, on the other hand, Stonehenge is ideal human-made memorial.

\section{Suggestions on the injection of  BDMs from SMBH}

 How the Black Hole can throw the matter out itself? There are huge relativistic jets that are known and visible phenomena of injection valuable masses from SMBH.  We see, that baryon-antibaryon matter lives in the torus shape. It just fill out the SMBH (what gravity without the matter?). Let us assume that this  matter can be injected in the similar BDM tori as it was absorbed in the Black Holes. The injection of supermassive BDM from SMBH is similar to its disintegration in the atmosphere: the result is descreate number of identical BDM tori with lower masses, which are getting accelerated due to released energy of initial BDM self connection. It seems, that massive torus has some critical condition, when small radius r begins equal to large one R with the growing mass of BH. The donut hole disappears, inner part of baryonium torus has to be rebuilded  into one or two BDM tori of lower mass. The best representation of critical torus seems the apple, see figure~ref{apple}. Since under the extra mass presure, the Black Matter of BH went to lower state with the emission a valuable part of its mass, radius of event horizon became smaller.  In such a way, the extra Dark Matter is situated already out of Black Hole event horizon and two BDM jets are getting accelerated due to the released potential energy of self baryonium connection. The extra baryonium matter is emitted in the directions of symmetry axe of torus as two back-to-back relativistic jets because of momentum conservation. At the end of BDM decay, baryoniums split step-by-step into ordinary protons and antiprotons. This subject of QCD string physics has multiple perspectives in the search for BDM debrieses: on one hand, such as observation of tetraquarks, pentaquarks, and hexaquarks on HE proton colliders, and, on the other hand, the study of Ultra High Energy (UHE) cosmic particle production in astrophysics.

\begin{figure}[h]
\begin{center}
\includegraphics[scale=0.26]{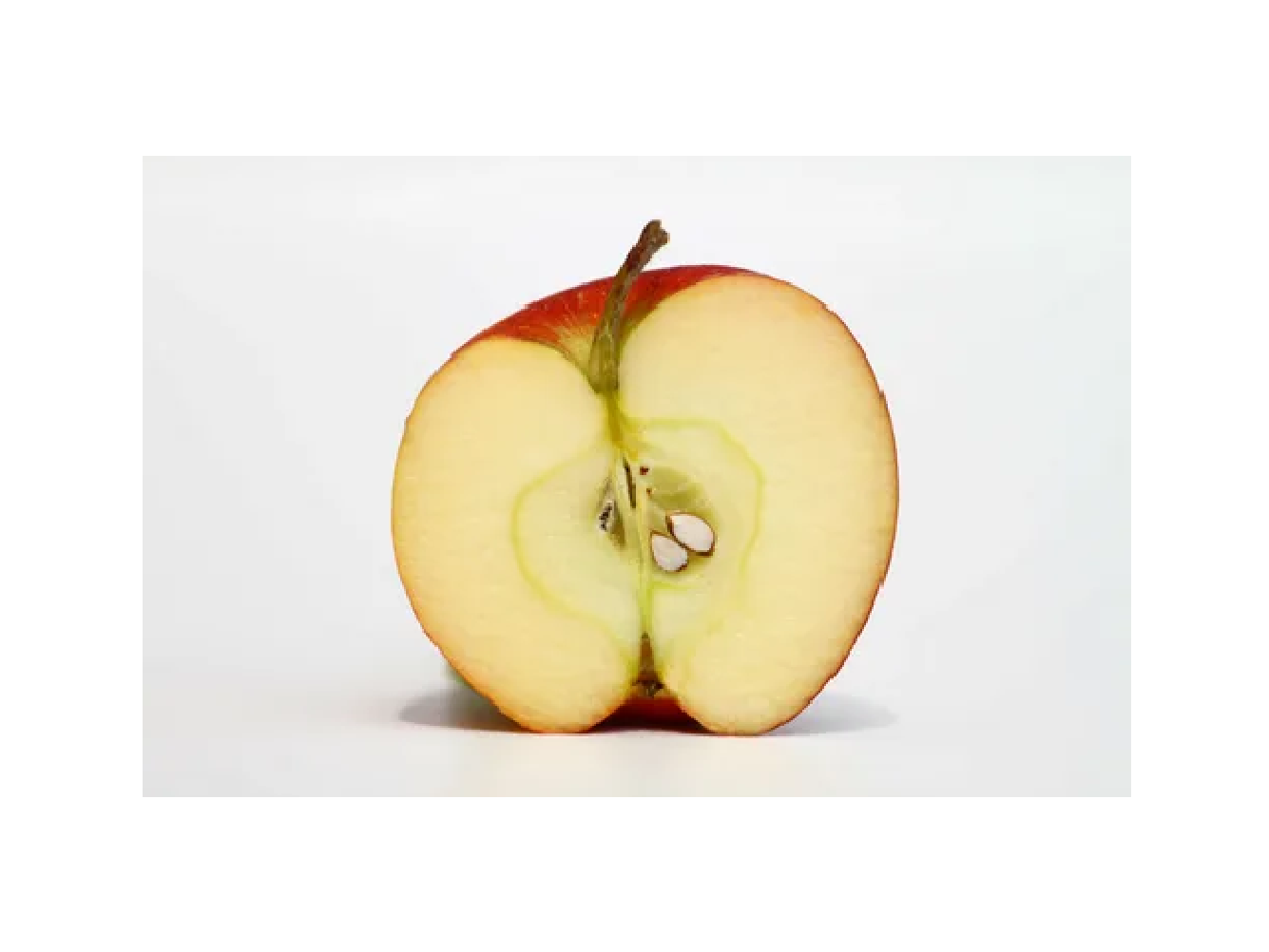}
\caption{Apple is a good example of degenerate torus.} 
\label{apple}
\end{center}
\end{figure}

\section{Why don't we meet BDM objects often?}

The BDM conglomerates are injected from the central SMBH of Milky Way approximately once in ten centuries  \cite{jetactivity}. They initially have mostly disintegrated in the process of jets injection, which is spreading baryon mass and energy throghout the Galaxy.
The last miserable share of super-massive BDMs can wander into the neghiborhood of Earth and originate the same processes as at relativistic jets: splitting and accelerating in atmosphere with the shower of various particles,the hydrogen is burning in oxigen atmosphere with water production, and, finally, BDM debrises hit the rocky surface and get absorbed in underground. What about dinosaurs, such catastrophic scenario can be realised once in the many tens of millions years at the tremendous activity of central SMBH in the Milky Way. If superheavy BDMs would be permanently coming from external source, we have been getting such disasters 330 times since 66 milions years ago, because our Galaxy is turning by 360 degrees once in 200 thousands years. Firs of all, the position of Earth should not be repeated in every cicle. And also the environement of our Galaxy is changing with time, so, may be the very massive BDM can not already reach the Solar system.

\section{Conclusions}
 
The ancient pits, wells, cenotes and othes holes that are distributed as circles in the undeground rocks are visible in many places on the surface of Earth. The circles of holes are the traces of disintergration of supermassive Baryonium Dark Matter. Since QCD matter obeys to certain symmetry, the neutral BDM state can fall apart only into the identical states with discrete number of equal lower masses, which have to be organized in the circles. The Baryonium Dark Matter objects consist of ordinary baryon-antibaryon  matter and may  possess the value of mass in the range from nucleon mass to the mass of order Black Hole. Such matter can be built only under very high pressure in Super Massive Black Holes or in Supernova explosions 
We observe on the Earth only heaviest parts of original BDM, because the light electromagnetic components are not leave the footprints. So these circles cannot be results of meteorit falls because of the absence of central crater. The first sign of heavy Baryonium Dark Matter with the mass above the known hadron masses was detected in 1975th in stratosphere experiment. Stratosphere event is real signature of BDM disintegration, because nothing can explain the circles in rapidity distribution at the region of intermediate rapidities and the scarce multiplicity of produced protons in forward area. Curcles are the results of disintegration of heavy mass BDM into the identical BDMs with lower masses. The likelihood exists to meet such objects in proximity of Earth. The stages of BDM disintegration in the atmosphere are similar to the BDM emission procedure with the relativistic jet: injection of  BDM in jet from SMBH, the splitting of supermassive BDM into lower-mass states, and the acceleration of secondaries due to the spare energy of SJ connections. As soon as rare super-massive BDM conglomerate reaches the atmosphere of  Earth, it splits apart and originates hadronic shower in atmosphere. The resulting heavy BDM debrises can punch the pits in the rocky ground. Since the components of torus are actually in a half the hydrogen nuclei, they are burning in oxigen atmosphere with blue flame and bring the water to the Earth. The important question for futher studies arises: how did the oxigen atmosphere appear at the Earth? The giant-BDM scenario, when dinosaurs went extinct, has not been yet realized again during few tens of million years. May be it cannot happen ever because the environment in Galaxy has been changed (more stars and dust don't allow BDM to penetrate through up to our Solar system). On the planets without atmosphere, the fall of super-massive BDM would cause the marsquakes.

\section{Acknowledgements} 
I am thankful to late Konstantine Aleksandrovich Kotelnikov for the fruitful discussions in the beginning of this work. He was the enthusiast of science and chief organizer of stratosphere experiment. I have also to thank the scientists of Theoretical Division of ITEP, where the knowledge on HE QCD interactions have been developed and retained. So why, I may name the Baryonium Dark Matter construction also as the compactificated Pomeron string.

\end{document}